\documentclass[sigconf]{acmart}

\usepackage{booktabs}
\usepackage{enumitem}
\usepackage{multirow}

\setcopyright{acmcopyright}

\acmDOI{10.1145/3132847.3133141}

\acmISBN{978-1-4503-4918-5/17/11}

\acmConference{CIKM'17 }{November 6--10, 2017}{Singapore, Singapore}
\acmYear{2017}
\copyrightyear{2017}

\acmPrice{15.00}

\begin{document}
\title{Extracting Entities of Interest from Comparative Product Reviews}

\author{Jatin Arora}
\affiliation{%
  \institution{Indian Institute of Technology Kharagpur}
}
\email{jatinarora2702@gmail.com}

\author{Sumit Agrawal}
\affiliation{%
  \institution{Indian Institute of Technology Kharagpur}
}
\email{agrawal.sumit33@gmail.com}

\author{Pawan Goyal}
\affiliation{%
  \institution{Indian Institute of Technology Kharagpur}
}
\email{pawang.iitk@gmail.com}

\author{Sayan Pathak}
\affiliation{%
  \institution{Microsoft Research, Redmond}
}
\email{sayanpa@microsoft.com}

\begin{abstract}
This paper presents a deep learning based approach to extract product comparison information out of user reviews on various e-commerce websites. Any comparative product review has three major entities of information: the names of the products being compared, the user opinion (predicate) and the feature or aspect under comparison. All these informing entities are dependent on each other and bound by the rules of the language, in the review. We observe that their inter-dependencies can be captured well using LSTMs. We evaluate our system on existing manually labeled datasets and observe out-performance over the existing Semantic Role Labeling (SRL) framework popular for this task.   
\end{abstract}


\maketitle

\section{Introduction}
User opinions have always had a strong influence on both producers and consumers in a market. In the past few years, with the advancement of e-commerce, a large proportion of these user opinions are present in the form of product reviews on online shopping websites like Amazon\footnote{https://www.amazon.com}, Ebay\footnote{http://www.ebay.com} etc. Product specifications bring out only the quantitative aspects of the product, but consumers are often interested in the qualitative comparison among competing products. Manufacturers, on the other hand, read product reviews to know the market response for their products and top competitors currently in the market. But going through the large volume of reviews manually has become increasingly difficult. Hence, automated extraction of this product comparison information from raw reviews is a popular research area.

\begin{figure*}
\includegraphics[width=0.50\textwidth]{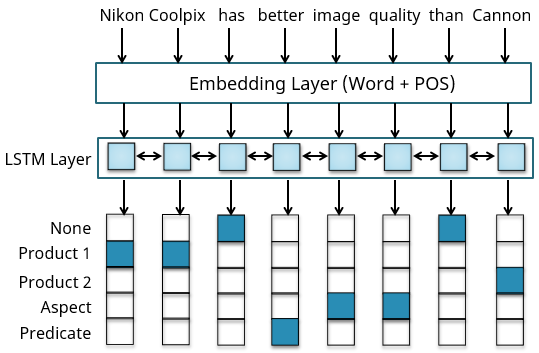}
\caption{Proposed model for extracting information from an example review}
\label{modelfig}
\vspace{-0.7em}
\end{figure*}

There can be various use cases for extracting information depending upon which, there can be variety of techniques to do the task. One can apply Named Entity Recognition (NER) to identify the products being compared and then do sentiment analysis to find the favored product. Sikchi et al. \cite{sikchi2016peq} use product specifications along with the review text for identifying the favored entity. Another way is to use text summarization to reduce the amount of text one has to manually read to infer the user opinion. Such techniques either do partial information extraction or require some manual intervention. We are interested in an automated full-scale extraction of comparison information from the reviews. In any review sentence involving comparison, there can be at most three major informing entities: the names of the two products being compared, the predicate or the user's opinion and the feature (aspect) under comparison. Consider an example camera review given by a user, "Nikon Coolpix has better image quality than Cannon". Given this review as input, we want to develop a system which can identify the products ("Nikon Coolpix", "Cannon"), the aspect being compared ("image quality") and the predicate or user opinion ("better"). A graphical representation of our system handling this example review is shown in Figure \ref{modelfig}.

Kessler and Kuhn \cite{kessler2013detection} model this as a Semantic Role Labeling (SRL) problem. In SRL, an event is expressed by the predicate (user opinion) and participants are the arguments that fill different semantic roles for the event. Here, the roles are the names of the products and the aspect being compared. They train a standard feature engineered SRL system \cite{bjorkelund2009multilingual} and show the best that can be achieved through it without  major adaptations.

We observe that all these informing entities (predicate, aspect and product names) are dependent on each other and extraction of one is facilitated by the knowledge of the other entities. This motivates us to model the sentence as a whole, using Long Short Term Memory (LSTM) cells, which inherently capture the inter-dependencies among these informing entities. Through this work, we show how combined extraction of informing entities using deep learning outperforms the existing feature-engineered frameworks. We compare with two SRL baselines and evaluate the systems on two tasks, argument identification and argument classification, and obtain better F1-Scores in both the tasks.

\section{Dataset}
Full-scale annotation of informing entities in a comparison sentence is a difficult task due to the diversity of writing styles of the users. So, most existing annotated datasets in this domain are small and manually labeled. Since a deep learning framework generally requires training through a large number of samples, we combine the annotated data obtained from various existing sources and split it (60:40) for training and testing. In addition to this, we artificially annotate review sentences using a pattern matching technique (explained in the next section) and add these to the training set. We, then filter out and use only review sentences which have at-least one comparative predicate and have length less than or equal to 30. The manually labeled datasets used are explained below.

\begin{itemize}[leftmargin=*]
\item \textbf{Jindal and Liu Corpus:} The corpus\footnote{Can be downloaded here, https://www.cs.uic.edu/\textasciitilde liub/FBS/data.tar.gz} contains review sentences mostly of products in electronics domain, annotated and segregated into 4 comparison categories. This was used by Jindal and Liu \cite{jindal2006mining, jindal2006identifying}. We use all comparison sentences from the corpus except type 4 (non-gradable comparisons). Each comparison sentence is annotated with names of the products (Entity 1 and 2), the aspect (Entity 3) and the predicate is mentioned as a bracketed comparison phrase.
\item \textbf{Corpus by Kessler and Kuhn:} This corpus \cite{kessler2014corpus} contains around 2200 manually annotated camera reviews. We use all the annotated sentences from here. The annotation scheme is the same as the one we use. Entities 1 and 2 are called products 1 and 2 in our nomenclature.
\item \textbf{JDPA Corpus:} This corpus \cite{KesslerEtAl2010} contains annotated blog posts containing user opinions about automobiles and digital cameras. We use only the sentences from the digital cameras domain which have the comparison class label in their annotation. The words marked by this class label bring out the user opinion and are marked as predicates. In addition, this class has 4 annotation slots, `More', `Less', `Dimension' and `Same'. We map the `More' slot to Product 1, `Less' slot to Product 2, `Dimension' slot to Aspect and ignore the `Same' slot which indicates if the two products are ranked as equal.
\item \textbf{Self Manual Labeling:} To include latest review trends, we crawled digital camera reviews from Amazon\footnote{https://www.amazon.com}, for the year 2016. Then, we manually annotated 350 review sentences with the three entities of information, wherever available.
\end{itemize}

Overall contribution of different corpora in our training and test data is summarized in Table~\ref{datasettable}.
\begin{table}[ht]
	\centering
\begin{tabular}{ | c || c | c | r | }
	\hline			
	Dataset & Train-Set & Test-Set & Total\\ \hline
	J\&L & 313 & 208 & 521\\ \hline
	Kessler & 982 & 655 & 1637\\ \hline
	JDPA & 133 & 90 & 223\\ \hline
    Manual & 210 & 140 & 350\\ \hline
    Pattern-Based & 24164 & 0 & 24164\\ \hline
    \textbf{Total} & \textbf{25802} & \textbf{1093} & \textbf{26895}\\ \hline
\end{tabular}
\caption{Datasets used in this study along with train-test split details}
\label{datasettable}
\vspace{-3em}
\end{table}

\section{Proposed Approach}

\subsection{Generation of Labeled Data}
We observe that there are some distinct styles for expressing comparison in product reviews, generally used by people. Based on this observation, we made 5 simple patterns using regular expressions. If an unlabeled review sentence matches a pattern, we narrow down the exact regions to look for different entities of information, based on the pattern. The predicate is then identified by a comparative POS tag (JJR, JJS, RBR, RBS - as per the Penn Treebank Tagging scheme). The aspect and product names are identified by dictionary matching. The aspects dictionary has 83 features for products in the electronics domain. The products dictionary is 11,126 entries long. Both these dictionaries are made semi-automatically, i.e., first using some heuristics to get a list with good accuracy and then manually correcting it. As an example, consider the pattern, [Aspect] [Preposition (\emph{of}|\emph{in})] [Product Name] [Opinion]. This pattern fits sentences like, "The zoom in Nikon S8100 is far better." and labels \emph{zoom} (Aspect), \emph{Nikon S8100} (Product1) and \emph{better} (Predicate). These patterns certainly do not exhaustively capture all possible comparisons, which is the final goal of this research work, but still give an annotated dataset with good precision, which can be used for training. We use this pattern fitting approach on electronic gadget reviews \cite{mcauley2013hidden} from Amazon\footnote{http://snap.stanford.edu/data/web-Amazon-links.html}. The labeled data hence generated is used in training only, as shown in Table~\ref{datasettable}. 

\subsection{Overall Framework}
Our model consists of three layers. An input review sentence is first tokenized and then its words are embedded by passing through the embedding layer. The embedded sentence is then passed through a LSTM (Long Short Term Memory) layer, where corresponding to each word, we have one LSTM unit. For each word of the sentence, the output from the corresponding LSTM cell is converted to a 5-dimensional attribute vector by passing through a fully connected layer. The attribute vector has one dimension for each entity of information (Product1, Product2, Aspect, Predicate, None) and is converted to a probability distribution by passing through a softmax layer. Finally, we take the label for the word/token as the attribute having the maximum probability. An example review being processed by our model is shown pictorially in Figure~\ref{modelfig}.

\subsection{Embeddings}
For a word/token in a sentence, the embedding layer finds out two embeddings, the word embedding and the one-hot POS (Part of Speech) embedding and concatenates the two, to be fed to the LSTM layer. We use the universal POS tags for POS embedding. For word embeddings, we try out 100-dimensional, and the standard 300-dimensional GloVe \cite{pennington2014glove} word embeddings trained on a general English corpus (Text8 Corpus\footnote{Can be downloaded from here, http://mattmahoney.net/dc/textdata}) and those trained specifically on electronics reviews from Amazon. We do not go for higher dimensional embeddings since that would increase the number of training parameters in our model and we may not be able to effectively train it using the current size of training data we have.

\subsection{Training and Model Variants}
We train our system to minimize the cross-entropy loss between the output probability distribution and the one-hot gold labels for tokens in sentences from the training set. There are several model variants that we test. We try out both unidirectional and bidirectional LSTMs. We work with both single and multiple LSTM layers. The specifications of all variants are shown in Table~\ref{modelspecstable} and the results obtained by these variations are all reported in the next section. The model giving the best results is shown in bold (Model2).

\begin{table}[ht]
\centering
\begin{tabular}{ | c || p{18mm} | p{9mm} | p{14.5mm} | p{14.5mm} | }
	\hline			
	Model & LSTM Type & LSTM Layers & Embedding Dimension & Embedding Source \\ \hline
	Model1 & Unidirectional & 1 & 300 & Text8\\ \hline
	\textbf{Model2} & \textbf{Bidirectional} & \textbf{1} & \textbf{300} & \textbf{Text8}\\ \hline
	Model3 & Unidirectional & 2 & 300 & Text8\\ \hline
	Model4 & Unidirectional & 1 & 100 & Text8\\ \hline
    Model5 & Unidirectional & 1 & 100 & Electronics\\ \hline
\end{tabular}
\caption{Specifications of the model variants, used in this study}
\label{modelspecstable}
\vspace{-1.5em}
\end{table}

\section{Experimental Results}

\subsection{Experimental Setup}
For sentence and word tokenization as well as POS Tagging, we use Natural Language Toolkit (NLTK). The deep learning model implementation is done using Tensorflow. The embeddings are prepared using GloVe and are kept frozen, not trained with the main model. All the parameters of the model are randomly initialized. For baseline approaches, using Semantic Role Labeling (SRL), we use the same settings as used by Kessler and Kuhn \cite{kessler2013detection}. The SRL system takes as input, data in CoNLL format for which we use the MATE\footnote{https://code.google.com/archive/p/mate-tools/} Dependency Parser \cite{bohnet2010very}.

\subsection{Evaluation Framework}
We test our system as well as the baselines, using the manually labeled test data described in Table~\ref{datasettable}. We evaluate the systems on two tasks and in both cases, we calculate the Precision, Recall and F1-Scores. The first task is argument identification i.e. identifying if a word/token has \emph{some} entity of information. The second task is, argument classification, where for a given word, the system has to classify it with one of the 5 labels (Predicate, Product 1, Product 2, Aspect, None).

\subsection{Baseline Approaches}
We compare our system with the approach presented in the paper by Kessler and Kuhn \cite{kessler2013detection}. The SRL is a feature engineered machine learning based system. Their system uses standard SRL features for extracting all informing entities in a review using a 2-stage pipeline. It first identifies only the predicate using SRL. Then, in the second stage, uses predicate information (either gold labeled predicates, or those identified in the first stage) for identifying and classifying the other arguments. In their paper, the authors present their results using gold predicates and report a 10\% decrease in the results if system identified predicates are used instead. We replicate their system and for a fair comparison with the proposed approach which is a single-stage model, we create two baselines. \textbf{Baseline1:} We use their method with the gold predicates information, and as mentioned in their paper, the results obtained from their system are decreased by 10\% to compare with the proposed model. \textbf{Baseline2:} Instead of gold predicates, we feed in the system identified predicates from stage 1 of the pipeline to stage 2 of the SRL system and compare with our model's performance. The results for predicate identification, argument identification and classification are shown in Tables~\ref{predidentifytable}, \ref{argidentifytable} and \ref{argclassifytable} respectively. Note that we do not show Baseline1 results in Table ~\ref{predidentifytable} as the gold standard predicates were used.

\begin{table*}[t]
\centering
\parbox{.45\linewidth}{
\centering
\begin{tabular}{ | c || c | c | r | }
\hline
Approach & Precision(\%) & Recall(\%) & F1-Score\\ \hline
Baseline2 & \textbf{82.4} & 3.4 & 6.5\\ \hline
Model1 & 72.0 & 25.4 & 37.6\\ \hline
\textbf{Model2} & 63.5 & \textbf{41.7} & \textbf{50.4}\\ \hline
Model3 & 47.5 & 18.1 & 26.2\\ \hline
Model4 & 66.2 & 20.9 & 31.8\\ \hline
Model5 & 69.7 & 28.3 & 40.2\\ \hline
\end{tabular}
\caption{Predicate Identification}
\label{predidentifytable}
}
\parbox{.45\linewidth}{
\centering
\begin{tabular}{ | c || c | c | r | }
\hline
Approach & Precision(\%) & Recall(\%) & F1-Score\\ \hline
Baseline1 & 62.2 & 30.6 & 41.0\\ \hline
Baseline2 & \textbf{67.9} & 1.7 & 3.3\\ \hline
Model1 & 67.8 & 21.3 & 28.8\\ \hline
\textbf{Model2} & 66.3 & \textbf{37.5} & \textbf{47.9}\\ \hline
Model3 & 66.1 & 13.6 & 17.9\\ \hline
Model4 & 64.1 & 15.8 & 20.2\\ \hline
Model5 & 67.0 & 13.8 & 18.6\\ \hline
\end{tabular}
\caption{Argument Identification} 
\label{argidentifytable}
}
\vspace{-2.2em}
\end{table*}
\begin{table*}[t]
	\centering
\begin{tabular}{ | c || c | c | c || c | c | c || c | c | r | }
	\hline			
	\multirow{2}{*}{Approach} & \multicolumn{3}{c ||}{Product 1} & \multicolumn{3}{c ||}{Product 2} & \multicolumn{3}{c |}{Aspect}\\ \cline{2-10}
     & Precision(\%) & Recall(\%) & F1-Score & Precision(\%) & Recall(\%) & F1-Score & Precision(\%) & Recall(\%) & F1-Score\\ \hline
	Baseline1 & 49.6 & 31.0 & 38.1 & 47.1 & 23.8 & 31.6 & 49.6 & 14.6 & 22.6\\ \hline
	Baseline2 & 54.1 & 1.8 & 3.5 & 55.0 & 1.5 & 2.9 & 45.5 & 0.4 & 0.8\\ \hline
    Model1 & 53.5 & 24.2 & 33.3 & 62.1 & 16.0 & 25.4 & \textbf{53.1} & 5.0 & 9.2\\ \hline
	\textbf{Model2} & 52.0 & \textbf{35.1} & \textbf{41.9} & 58.8 & \textbf{30.6} & \textbf{40.3} & 46.8 & \textbf{21.6} & \textbf{29.3}\\ \hline
    Model3 & 53.3 & 12.0 & 19.6 & 57.8 & 10.5 & 17.8 & 21.1 & 0.3 & 0.6\\ \hline
    Model4 & 54.4 & 19.3 & 28.6 & 60.4 & 12.1 & 20.2 & 51.7 & 2.5 & 4.9\\ \hline
    Model5 & \textbf{58.3} & 17.6 & 27.0 & \textbf{66.3} & 8.0 & 14.3 & 46.4 & 2.7 & 5.0\\ \hline
\end{tabular}
\caption{Argument Classification}
\label{argclassifytable}
\vspace{-3em}
\end{table*}

We observe that a single layer of Bidirectional LSTMs, using 300 dimensional GloVe word embeddings prepared from general English (Text8) Corpus gives the best results overall and outperforms both baselines in all the tasks in terms of recall as well as F1-score\footnote{This corresponds to Model2, shown in Bold.}.

\section{Discussions}
\begin{itemize}[leftmargin=*]
\item Since a large amount of training data is generated using patterns, we observe a relatively low recall from the models trained using the data, as expected. But Baseline2 reports a very low recall. This is because, in the pipelined SRL approach, correct identification of the predicate (the event) is key to further identification of arguments (roles). Since Baseline2 gives a high precision and very low recall for predicate identification itself on the test data, hence same is the trend for argument identification and classification as well. Our system, on the other hand, overcomes the limitation of a pipelined approach by combined modeling of the informing entities.

\item Increasing the number of hidden LSTM layers does not improve the results, thus confirming that a single layer LSTM rightly captures the dependencies among the informing entities in a comparison based review sentence.

\item Using 100 dimensional word embeddings leads to a lower recall. But, since the embedding dimensions are proportional to trainable model parameters, smaller dimensional embeddings can give a good enough model even when the training set is small.

\item Embeddings specifically prepared from the electronics corpus give a slightly better precision but compromise with the recall. Hence, general English text embeddings and electronics embeddings both give almost similar F1-score on both tasks.

\vspace{-0.5em}
\end{itemize}

\section{Conclusions and Future Work}
In this paper, we presented a simple framework which uses deep learning to annotate and hence, extract all important entities of information from comparative product reviews. This system saves the trouble of feature engineering and gives better results than the previously presented SRL based system.

We also developed simple patterns which capture some common styles of presenting comparisons in reviews. This pattern fitting technique proved beneficial in expanding our training data, making it possible for the deep learning model to effectively learn the sentential structure and inter-dependencies among the informing entities in comparative reviews.

There is still a lot of scope for improvement. In reviews, users often tend to use pronouns or refer implicitly to a product mentioned in the previous sentences. In such cases, a wrapper system needs to be developed which can capture the sentence-to-sentence dependencies and map the pronoun in the current sentence, to the corresponding noun mentioned in the previous sentences. This is an active area of research which we would like to explore. Peng et al. \cite{peng2017cross} show the effectiveness of Graph LSTMs for such cross-sentence relation extraction.
\vspace{-0.5em}

\begin{acks}
The authors would like to acknowledge the contributions of Yash Agrawal and Kushagra Aggarwal, CSE, IIT Kharagpur, in parsing the JDPA Corpus and manual annotation of datasets.
\end{acks}

\vspace{-0.3em}
\bibliographystyle{ACM-Reference-Format}
\bibliography{references} 

\end{document}